%% file: main.tex
\pdfoutput=1

\documentclass[12pt,a4paper]{article}

\usepackage{ifthen} 
\newboolean{pdflatex}
\setboolean{pdflatex}{true} 

\newboolean{articletitles}
\setboolean{articletitles}{true} 

\newboolean{uprightparticles}
\setboolean{uprightparticles}{false} 

\newboolean{inbibliography}
\setboolean{inbibliography}{false} 

\input{preamble}
\usepackage{relsize}

\begin{document}

\renewcommand{\thefootnote}{\fnsymbol{footnote}}
\setcounter{footnote}{1}

\input{title-LHCb-PAPER}

\renewcommand{\thefootnote}{\arabic{footnote}}
\setcounter{footnote}{0}



\pagestyle{plain} 
\setcounter{page}{1}
\pagenumbering{arabic}


%

\input{introduction}

\input{detector}

\input{selection}

\input{fit}

\input{systematics}

\input{bfresult}

\input{acp}

\input{conclusions}

\input{acknowledgements}



\addcontentsline{toc}{section}{References}
\setboolean{inbibliography}{true}
\bibliographystyle{LHCb}
\bibliography{main,references,LHCb-PAPER,LHCb-CONF,LHCb-DP,LHCb-TDR}
\newpage


\clearpage

\input{LHCb_HD_authorlist_2016-01-19}

\end{document}

%% file: preamble.tex
\usepackage{microtype}
\usepackage{lineno}  
\usepackage{xspace} 
\usepackage{caption} 

\usepackage{graphicx}  
\usepackage{color}
\usepackage{colortbl}
\graphicspath{{./figs/}} 

\usepackage{amsmath} 
\usepackage{amssymb}
\usepackage{amsfonts}
\usepackage{upgreek} 

\newcommand*\patchAmsMathEnvironmentForLineno[1]{%
\expandafter\let\csname old#1\expandafter\endcsname\csname #1\endcsname
\expandafter\let\csname oldend#1\expandafter\endcsname\csname
end#1\endcsname
 \renewenvironment{#1}%
   {\linenomath\csname old#1\endcsname}%
   {\csname oldend#1\endcsname\endlinenomath}%
}
\newcommand*\patchBothAmsMathEnvironmentsForLineno[1]{%
  \patchAmsMathEnvironmentForLineno{#1}%
  \patchAmsMathEnvironmentForLineno{#1*}%
}
\AtBeginDocument{%
\patchBothAmsMathEnvironmentsForLineno{equation}%
\patchBothAmsMathEnvironmentsForLineno{align}%
\patchBothAmsMathEnvironmentsForLineno{flalign}%
\patchBothAmsMathEnvironmentsForLineno{alignat}%
\patchBothAmsMathEnvironmentsForLineno{gather}%
\patchBothAmsMathEnvironmentsForLineno{multline}%
\patchBothAmsMathEnvironmentsForLineno{eqnarray}%
}

\usepackage{hyperref}    
\usepackage[all]{hypcap} 

\input{lhcb-symbols-def} 

\usepackage{cite} 
\usepackage{mciteplus}

%% file: lhcb-symbols-def.tex
\def\lhcb {\mbox{LHCb}\xspace}

\def\MagUp {\mbox{\em Mag\kern -0.05em Up}\xspace}

\ifthenelse{\boolean{uprightparticles}}%
{

 \def\Pmu         {\ensuremath{\upmu}\xspace}

 \def\Ppi         {\ensuremath{\uppi}\xspace}

 \def\Ppsi        {\ensuremath{\uppsi}\xspace}

 \def\PDelta      {\ensuremath{\Delta}\xspace}
 \def\PXi      {\ensuremath{\Xi}\xspace}
 \def\PLambda      {\ensuremath{\Lambda}\xspace}
 \def\PSigma      {\ensuremath{\Sigma}\xspace}
 \def\POmega      {\ensuremath{\Omega}\xspace}
 \def\PUpsilon      {\ensuremath{\Upsilon}\xspace}

 \def\PB      {\ensuremath{\mathrm{B}}\xspace}
 
 \def\PD      {\ensuremath{\mathrm{D}}\xspace}

 \def\PJ      {\ensuremath{\mathrm{J}}\xspace}
 \def\PK      {\ensuremath{\mathrm{K}}\xspace}

 \def\Pb      {\ensuremath{\mathrm{b}}\xspace}
 \def\Pc      {\ensuremath{\mathrm{c}}\xspace}

 \def\Pi      {\ensuremath{\mathrm{i}}\xspace}

 \def\Pp      {\ensuremath{\mathrm{p}}\xspace}

}
{

 \def\Pmu         {\ensuremath{\mu}\xspace}

 \def\Ppi         {\ensuremath{\pi}\xspace}

 \def\Ppsi        {\ensuremath{\psi}\xspace}
 
 \mathchardef\PDelta="7101
 \mathchardef\PXi="7104
 \mathchardef\PLambda="7103
 \mathchardef\PSigma="7106
 \mathchardef\POmega="710A
 \mathchardef\PUpsilon="7107
 
 \def\PB      {\ensuremath{B}\xspace}
 
 \def\PD      {\ensuremath{D}\xspace}

 \def\PJ      {\ensuremath{J}\xspace}
 \def\PK      {\ensuremath{K}\xspace}

 \def\Pb      {\ensuremath{b}\xspace}
 \def\Pc      {\ensuremath{c}\xspace}

 \def\Pi      {\ensuremath{i}\xspace}

 \def\Pp      {\ensuremath{p}\xspace}

}

\makeatletter
\ifcase \@ptsize \relax
  \newcommand{\miniscule}{\@setfontsize\miniscule{4}{5}}
\or
  \newcommand{\miniscule}{\@setfontsize\miniscule{5}{6}}
\or
  \newcommand{\miniscule}{\@setfontsize\miniscule{5}{6}}
\fi
\makeatother

\DeclareRobustCommand{\optbar}[1]{\shortstack{{\miniscule (\rule[.5ex]{1.25em}{.18mm})}
  \\ [-.7ex] $#1$}}

\def\mumu       {{\ensuremath{\Pmu^+\Pmu^-}}\xspace}

\def\cquark    {{\ensuremath{\Pc}}\xspace}

\def\bquark    {{\ensuremath{\Pb}}\xspace}

\def\pion   {{\ensuremath{\Ppi}}\xspace}
\def\piz    {{\ensuremath{\pion^0}}\xspace}

\def\pip    {{\ensuremath{\pion^+}}\xspace}
\def\pim    {{\ensuremath{\pion^-}}\xspace}

\def\pimp   {{\ensuremath{\pion^\mp}}\xspace}

\def\kaon    {{\ensuremath{\PK}}\xspace}
  \def\Kbar    {{\kern 0.2em\overline{\kern -0.2em \PK}{}}\xspace}

\def\KorKbar    {\kern 0.18em\optbar{\kern -0.18em K}{}\xspace}

\def\Kp      {{\ensuremath{\kaon^+}}\xspace}
\def\Km      {{\ensuremath{\kaon^-}}\xspace}
\def\Kpm     {{\ensuremath{\kaon^\pm}}\xspace}

\def\KS      {{\ensuremath{\kaon^0_{\rm\scriptscriptstyle S}}}\xspace}

\def\Kstarp  {{\ensuremath{\kaon^{*+}}}\xspace}

  \def\Dbar    {{\kern 0.2em\overline{\kern -0.2em \PD}{}}\xspace}
\def\D       {{\ensuremath{\PD}}\xspace}

\def\DorDbar    {\kern 0.18em\optbar{\kern -0.18em D}{}\xspace}
\def\Dz      {{\ensuremath{\D^0}}\xspace}

\def\Bbar    {{\ensuremath{\kern 0.18em\overline{\kern -0.18em \PB}{}}}\xspace}

\def\BorBbar    {\kern 0.18em\optbar{\kern -0.18em B}{}\xspace}

\def\jpsi     {{\ensuremath{{\PJ\mskip -3mu/\mskip -2mu\Ppsi\mskip 2mu}}}\xspace}

  \def\Y#1S{\ensuremath{\PUpsilon{(#1S)}}\xspace}

\def\proton      {{\ensuremath{\Pp}}\xspace}

\def\Xires       {{\ensuremath{\PXi}}\xspace}

\def\Lz          {{\ensuremath{\PLambda}}\xspace}
\def\Lbar        {{\ensuremath{\kern 0.1em\overline{\kern -0.1em\PLambda}}}\xspace}
\def\LorLbar    {\kern 0.18em\optbar{\kern -0.18em \PLambda}{}\xspace}

\def\Lb      {{\ensuremath{\Lz^0_\bquark}}\xspace}
\def\Lbbar   {{\ensuremath{\Lbar{}^0_\bquark}}\xspace}
\def\Lc      {{\ensuremath{\Lz^+_\cquark}}\xspace}

\def\Xib     {{\ensuremath{\Xires_\bquark}}\xspace}
\def\Xibz    {{\ensuremath{\Xires^0_\bquark}}\xspace}
\def\Xibm    {{\ensuremath{\Xires^-_\bquark}}\xspace}

\def\Xicp    {{\ensuremath{\Xires^+_\cquark}}\xspace}

\def\to                 {\ensuremath{\rightarrow}\xspace}

\def\CP                {{\ensuremath{C\!P}}\xspace}

\def\AT#1     {\ensuremath{A_{\mathrm{T}}^{#1}}\xspace}           

\def\C#1      {\ensuremath{\mathcal{C}_{#1}}\xspace}                       
\def\Cp#1     {\ensuremath{\mathcal{C}_{#1}^{'}}\xspace}                    
\def\Ceff#1   {\ensuremath{\mathcal{C}_{#1}^{\mathrm{(eff)}}}\xspace}        
\def\Cpeff#1  {\ensuremath{\mathcal{C}_{#1}^{'\mathrm{(eff)}}}\xspace}       
\def\Ope#1    {\ensuremath{\mathcal{O}_{#1}}\xspace}                       
\def\Opep#1   {\ensuremath{\mathcal{O}_{#1}^{'}}\xspace}                    

\newcommand{\tev}{\ifthenelse{\boolean{inbibliography}}{\ensuremath{~T\kern -0.05em eV}\xspace}{\ensuremath{\mathrm{\,Te\kern -0.1em V}}}\xspace}
\newcommand{\gev}{\ensuremath{\mathrm{\,Ge\kern -0.1em V}}\xspace}
\newcommand{\mev}{\ensuremath{\mathrm{\,Me\kern -0.1em V}}\xspace}
\newcommand{\kev}{\ensuremath{\mathrm{\,ke\kern -0.1em V}}\xspace}
\newcommand{\ev}{\ensuremath{\mathrm{\,e\kern -0.1em V}}\xspace}
\newcommand{\gevc}{\ensuremath{{\mathrm{\,Ge\kern -0.1em V\!/}c}}\xspace}
\newcommand{\mevc}{\ensuremath{{\mathrm{\,Me\kern -0.1em V\!/}c}}\xspace}
\newcommand{\gevcc}{\ensuremath{{\mathrm{\,Ge\kern -0.1em V\!/}c^2}}\xspace}
\newcommand{\gevgevcccc}{\ensuremath{{\mathrm{\,Ge\kern -0.1em V^2\!/}c^4}}\xspace}
\newcommand{\mevcc}{\ensuremath{{\mathrm{\,Me\kern -0.1em V\!/}c^2}}\xspace}

\def\mm   {\ensuremath{\rm \,mm}\xspace}

\def\mum  {\ensuremath{{\,\upmu\rm m}}\xspace}

\def\invfb   {\ensuremath{\mbox{\,fb}^{-1}}\xspace}

\newcommand{\chisq}{\ensuremath{\chi^2}\xspace}

\newcommand{\chisqip}{\ensuremath{\chi^2_{\rm IP}}\xspace}
\newcommand{\chisqvs}{\ensuremath{\chi^2_{\rm VS}}\xspace}
\newcommand{\chisqvtx}{\ensuremath{\chi^2_{\rm vtx}}\xspace}

\def\gsim{{~\raise.15em\hbox{$>$}\kern-.85em
          \lower.35em\hbox{$\sim$}~}\xspace}
\def\lsim{{~\raise.15em\hbox{$<$}\kern-.85em
          \lower.35em\hbox{$\sim$}~}\xspace}

\def\sPlot{\mbox{\em sPlot}\xspace}

\def\ptot       {\mbox{$p$}\xspace}
\def\pt         {\mbox{$p_{\rm T}$}\xspace}

\def\evtgen     {\mbox{\textsc{EvtGen}}\xspace}

\def\geant      {\mbox{\textsc{Geant4}}\xspace}

\def\photos     {\mbox{\textsc{Photos}}\xspace}

\def\pythia     {\mbox{\textsc{Pythia}}\xspace}

\def\tell1  {TELL1\xspace}
\def\ukl1   {UKL1\xspace}

\newcommand{\ie}{\mbox{\itshape i.e.}\xspace}

%% file: title-LHCb-PAPER.tex
\begin{titlepage}
\pagenumbering{roman}

\vspace*{-1.5cm}
\centerline{\large EUROPEAN ORGANIZATION FOR NUCLEAR RESEARCH (CERN)}
\vspace*{1.5cm}
\noindent
\begin{tabular*}{\linewidth}{lc@{\extracolsep{\fill}}r@{\extracolsep{0pt}}}
\ifthenelse{\boolean{pdflatex}}
{\vspace*{-2.7cm}\mbox{\!\!\!\includegraphics[width=.14\textwidth]{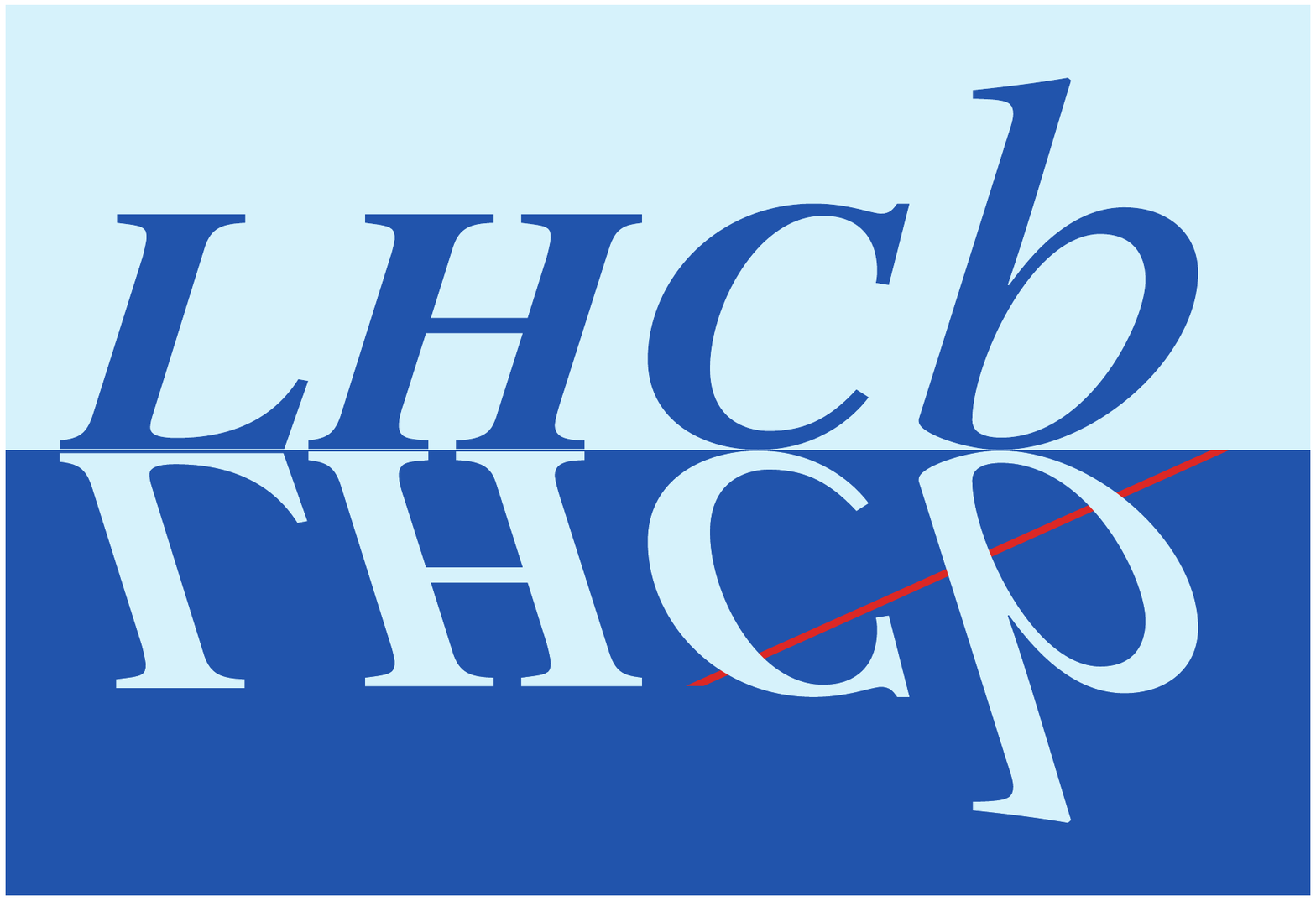}} & &}%
{\vspace*{-1.2cm}\mbox{\!\!\!\includegraphics[width=.12\textwidth]{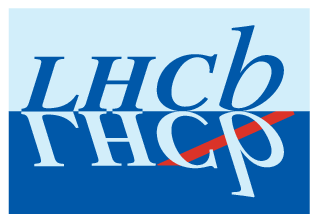}} & &}%
\\
 & & CERN-EP-2016-038 \\  
 & & LHCb-PAPER-2016-004 \\  
 & & \today \\ 
 & & \\
\end{tabular*}

\vspace*{2.0cm}

{\bf\boldmath\Large
\begin{center}
  Observations of $\Lb \to \Lz\Kp\pim$ and $\Lb \to \Lz\Kp\Km$ decays
  and searches for other $\Lb$ and $\Xibz$ decays to $\Lz h^+h^{\prime -}$ final states
\end{center}
}

\vspace*{1.0cm}

\begin{center}
The LHCb collaboration\footnote{Authors are listed at the end of this paper.}
\end{center}

\vspace{\fill}

\begin{abstract}
  \noindent
  A search is performed for the charmless three-body decays of the $\Lb$ and $\Xibz$ baryons to the final states $\Lz h^+h^{\prime -}$, where $h^{(\prime)} = \pi$ or $K$.
  The analysis is based on a data sample, corresponding to an integrated luminosity of $3\invfb$ of $pp$ collisions, collected by the LHCb experiment.
  The $\Lb \to \Lz\Kp\pim$ and $\Lb \to \Lz\Kp\Km$ decays are observed for the first time and their branching fractions and \CP asymmetry parameters are measured.
  Evidence is seen for the $\Lb \to \Lz\pip\pim$ decay and limits are set on the branching fractions of $\Xibz$ baryon decays to the  $\Lz h^+h^{\prime -}$ final states.
\end{abstract}

\vspace*{1.0cm}

\begin{center}
  Published in JHEP 05 (2016) 081 
\end{center}

\vspace{\fill}

{\footnotesize
\centerline{\copyright~CERN on behalf of the \lhcb collaboration, licence \href{http://creativecommons.org/licenses/by/4.0/}{CC-BY-4.0}.}}
\vspace*{2mm}

\end{titlepage}


\newpage
\setcounter{page}{2}
\mbox{~}
%
%
%
%

\cleardoublepage

%% file: introduction.tex
\section{Introduction}
\label{sec:Introduction}

The availability of large samples of high energy $pp$ collision data has allowed significant improvements in the experimental studies of $b$ baryons.
The masses and lifetimes of the $\Lb$, $\Xibz$ and $\Xibm$ particles are all now known to within a few percent or better~\cite{LHCb-PAPER-2014-002,LHCb-PAPER-2014-003,LHCb-PAPER-2014-010,LHCb-PAPER-2014-021,LHCb-PAPER-2014-048}, and excited \Lb and \Xib baryons have been discovered~\cite{LHCb-PAPER-2012-012,Chatrchyan:2012ni,LHCb-PAPER-2014-061}.
However, relatively few decay modes of the $b$ baryons have yet been studied.
In particular, among the possible charmless hadronic final states, only the two-body $\Lb \to \proton\Km$ and $\Lb \to \proton\pim$ decays~\cite{Aaltonen:2008hg}, the quasi-two-body $\Lb \to \Lz\phi$ decay~\cite{LHCb-PAPER-2016-002} and the three-body $\Lb \to \KS\proton\pim$ decay~\cite{LHCb-PAPER-2013-061} have been observed, while evidence has been reported for the $\Lb \to \Lz \eta$ decay~\cite{LHCb-PAPER-2015-019}.
No decay of a \Xib baryon to a charmless final state has yet been observed.
Such decays are of great interest as they proceed either by tree-level decays involving the Cabibbo-Kobayashi-Maskawa~\cite{Cabibbo:1963yz,Kobayashi:1973fv} matrix element $V^{}_{ub}$ or by loop-induced amplitudes, and they are consequently expected to have suppressed decay rates in the Standard Model.
Their study may also provide insights into the mechanisms of hadronisation in $b$ baryon decays.
Moreover, charmless hadronic $b$ baryon decays provide interesting possibilities to search for \CP violation effects, as have been seen in the corresponding $B$ meson decays~\cite{Lees:2012mma,Duh:2012ie,Aaltonen:2011qt,LHCb-PAPER-2013-018,LHCb-PAPER-2014-044}.

In this paper, a search is reported for charmless decays of the $\Lb$ and $\Xibz$ baryons to the final states $\Lz \pip\pim$, $\Lz \Kpm\pimp$ and $\Lz \Kp\Km$.
The inclusion of charge conjugate processes is implied throughout, except where the determination of asymmetries is discussed.
Intermediate states containing charmed hadrons are excluded from the signal sample and studied separately: transitions involving a $\Lc \to \Lz\pip$ decay are used as a control sample and to normalise the measured branching fractions, and those with $\Lc \to \Lz\Kp$ decays provide cross-checks of the analysis procedure.
In all cases the $\Lz$ baryon is reconstructed in the $\proton\pim$ final state.
Although \bquark baryon decays to the $\Lz\Kp\pim$ and $\Lz\Km\pip$ final states can be distinguished through correlation of the proton and kaon charges, they are combined together in the $\Lz \Kpm\pimp$ sample to improve the stability of the fit to the mass spectra.
The $\Lb\to\Lz\Kp\pim$ and $\Xibz\to\Lz\Km\pip$ decays are expected to dominate over the modes with swapped kaon and pion charges, and therefore the results are presented assuming the suppressed contribution is negligible, as is commonly done in similar cases~\cite{Duh:2012ie,Aubert:2006fha,Aaltonen:2011qt,LHCb-PAPER-2012-002}.

No previous experimental information exists on the charmless hadronic decays being studied; theoretical predictions for the branching fraction of the $\Lb \to \Lz\pip\pim$ decay are in the range $10^{-9}$--$10^{-7}$~\cite{Guo:1998eg,Arunagiri:2003gu,Leitner:2006nb}.

The paper is organised as follows.
A description of the LHCb detector and the dataset used for the analysis is given in Sec.~\ref{sec:Detector}.
The selection algorithms, the method to determine signal yields, and the systematic uncertainties on the results are discussed in Secs.~\ref{sec:Selection}--\ref{sec:Systematics}.
The measured branching fractions are presented in Sec.~\ref{sec:bfresults}.
Since significant signals are observed for the $\Lb \to \Lz\Kp\pim$ and $\Lb \to \Lz\Kp\Km$ channels, measurements of the phase-space integrated \CP asymmetry parameters of these modes are reported in Sec.~\ref{sec:acp}.
Conclusions are given in Sec.~\ref{sec:Conclusions}.

%% file: detector.tex
\section{Detector and dataset}
\label{sec:Detector}

The analysis is based on $pp$ collision data collected with the LHCb detector, corresponding to $1.0\invfb$ at a centre of mass energy of $7 \tev$ in 2011, and $2.0\invfb$ at a centre of mass energy of $8 \tev$ in 2012.
The \lhcb detector~\cite{Alves:2008zz,LHCb-DP-2014-002} is a single-arm forward spectrometer covering the \mbox{pseudorapidity} range $2<\eta <5$, designed for the study of particles containing \bquark or \cquark quarks.
The detector includes a high-precision tracking system consisting of a silicon-strip vertex detector surrounding the $pp$ interaction region, a large-area silicon-strip detector located upstream of a dipole magnet with a bending power of about $4{\rm\,Tm}$, and three stations of silicon-strip detectors and straw drift tubes placed downstream of the magnet.
The tracking system provides a measurement of momentum, \ptot, of charged particles with a relative uncertainty that varies from 0.5\% at low momentum to 1.0\% at 200\gevc.
The minimum distance of a track to a primary vertex, the impact parameter (IP), is measured with a resolution of $(15+29/\pt)\mum$, where \pt is the component of the momentum transverse to the beam, in \gevc.
Different types of charged hadrons are distinguished using information from two ring-imaging Cherenkov detectors.
Photons, electrons and hadrons are identified by a calorimeter system consisting of scintillating-pad and preshower detectors, an electromagnetic calorimeter and a hadronic calorimeter.
Muons are identified by a system composed of alternating layers of iron and multiwire proportional chambers.

The online event selection is performed by a trigger~\cite{LHCb-DP-2012-004,LHCb-PUB-2014-046}, which consists of a hardware stage, based on information from the calorimeter and muon systems, followed by a software stage, in which all charged particles with $\pt>500~(300)\mevc$ are reconstructed for 2011 (2012) data.
At the hardware trigger stage, events are required to have a muon with high \pt or a hadron, photon or electron with high transverse energy in the calorimeters.
For hadrons, the transverse energy threshold is 3.5\gev.
The software trigger requires a two-, three- or four-track secondary vertex with significant displacement from the primary $pp$ interaction vertices~(PVs).
At least one charged particle must have transverse momentum $\pt > 1.7\gevc$ and be inconsistent with originating from any PV.
A multivariate algorithm~\cite{BBDT} is used for the identification of secondary vertices consistent with the decay of a \bquark hadron.

The efficiency with which the software trigger selected the signal modes varied during the data-taking period, for reasons that are related to the reconstruction of the long-lived $\Lz$ baryon.
Such decays are reconstructed in two different categories, the first involving $\Lz$ particles that decay early enough for the produced particles to be reconstructed in the vertex detector, and the second containing $\Lz$ baryons that decay later such that track segments cannot be formed in the vertex detector.
These categories are referred to as \emph{long} and \emph{downstream}, respectively.
During 2011, downstream tracks were not reconstructed in the software trigger.
Such tracks were included in the trigger logic during 2012 data-taking; however, a significant improvement in the algorithms was implemented during a technical stop period.
Consequently, the data are subdivided into three data-taking periods (2011, 2012a and 2012b) as well as the two reconstruction categories (long and downstream).
The 2012b sample has the best trigger efficiency, especially in the downstream category, and is also the largest sample, corresponding to $1.4\invfb$.
The long category has better mass, momentum and vertex resolution than the downstream category.

Simulated data samples are used to study the response of the detector and to investigate certain categories of background.
In the simulation, $pp$ collisions are generated using \pythia~\cite{Sjostrand:2006za,*Sjostrand:2007gs} with a specific \lhcb configuration~\cite{LHCb-PROC-2010-056}.
Decays of hadronic particles are described by \evtgen~\cite{Lange:2001uf}, in which final-state radiation is generated using \photos~\cite{Golonka:2005pn}.
The interaction of the generated particles with the detector, and its response, are implemented using the \geant toolkit~\cite{Allison:2006ve, *Agostinelli:2002hh} as described in Ref.~\cite{LHCb-PROC-2011-006}.

%% file: selection.tex
\section{Selection requirements and efficiency modelling}
\label{sec:Selection}

The selection exploits the topology of the three-body decay and the \bquark baryon kinematic properties, first in a preselection stage, with minimal effect on signal efficiency, and subsequently in a multivariate classifier.
Each \bquark baryon candidate is reconstructed by combining two oppositely charged tracks with a \Lz candidate.
The \Lz decay products are both required to have momentum greater than $2\gevc$ and to form a vertex with low \chisqvtx.

In addition,
the tracks must not be associated with any PV as quantified by the \chisqip variable, defined as the difference in \chisq of a given PV reconstructed with and without the considered track.

The track pair must satisfy $|m(\proton\pim) - m_{\Lz}| < 20\,(15)\,\mevcc$ for downstream (long) candidates, where $m_{\Lz}$ is the known \Lz mass~\cite{PDG2014}.
The \Lz candidate is associated to the PV which gives the smallest \chisqip, and significant vertex separation is ensured with a requirement on \chisqvs, the square of the separation distance between the \Lz vertex and the associated PV divided by its uncertainty.
A loose particle identification (PID) requirement, based primarily on information from the ring-imaging Cherenkov detectors, is imposed on the proton candidate to remove background from $\KS$ decays.
For downstream \Lz candidates $p_{\Lz} > 8\gevc$ is also required.

The scalar sum of the transverse momenta of the \Lz candidate and the two $h^+h^{\prime -}$ tracks is required to be greater than $3\gevc$ ($4.2\gevc$ for downstream candidates).

The IP of the charged track with the largest \pt is required to be greater than $0.05 \mm$.
The minimum, for any pair from $(\Lz, h^+, h^{\prime -})$, of the square of the distance of closest approach divided by its uncertainty must be less than $5$.
The \bquark baryon candidate must have a good quality vertex, be significantly displaced from the PV, and have $\pt > 1.5\,\gevc$.
Furthermore, it must have low values of both \chisqip and pointing angle (\ie\ the angle between the \bquark baryon momentum vector and the line joining its production and decay vertices), which ensure that it points back to the PV.
Additionally, the \Lz and \bquark baryon candidate vertices must be separated by at least $30 \mm$ along the beam direction.
The candidates are separated with PID criteria (discussed below) into the three different final states: $\Lz\pip\pim$, $\Lz\Kpm\pimp$ and $\Lz\Kp\Km$.
Candidates where any of the tracks is identified as a muon are rejected; this removes backgrounds resulting from semimuonic \bquark baryon decays, $\jpsi \to \mumu$ decays, or $\Lb \to \Lz\mumu$ decays~\cite{LHCb-PAPER-2015-009}.
Decays involving intermediate \Lc baryons are removed from the signal sample with a veto that is applied within $\pm 30\mevcc$ of the known \Lc mass~\cite{PDG2014}; in the case of \Lc \to\Lz\pip however, these candidates are retained and used as a control sample.
A similar veto window is applied around the \Xicp mass, and backgrounds from the $\Lb \to \Lz \Dz$ decay with $\Dz \to h^+ h^{\prime -}$ are also removed with a $\pm 30\mevcc$ window around the known \Dz mass.

The \bquark baryon candidates are required to have invariant mass within the range $5300 < m(\Lz h^+ h^{\prime -}) < 6100 \mevcc$, when reconstructed with the appropriate mass hypothesis for the $h^+$ and $h^{\prime -}$ tracks.
To avoid potential biases during the selection optimisation, regions of $\pm 50\mevcc$, to be compared to the typical resolution of $15\mevcc$, around both the \Lb and \Xibz masses were not examined until the selection criteria were established.

Further separation of signal from combinatorial background candidates is achieved with a boosted decision tree (BDT) multivariate classifier~\cite{Breiman,AdaBoost}.
The BDT is trained using a simulated $\Lb \to \Lz\pip\pim$ signal sample and data from the sideband region $5838 < m(\Lz \pip\pim) < 6100 \mevcc$ for the background.
To prevent bias, each sample is split into two disjoint subsets and two separate classifiers are each trained and evaluated on different subsets, such that events used to train one BDT are classified using the other.

The set of input variables is chosen to optimise the performance of the algorithm, and to minimise variation of the efficiency across the phase space.
The input variables for the BDTs are: $\pt$, $\eta$, \chisqip, \chisqvs, pointing angle and \chisqvtx of the \bquark baryon candidate; the sum of the \chisqip values of the $h^{+}$ and $h^{\prime -}$ tracks; and the \chisqip, \chisqvs and \chisqvtx of the \Lz candidate.
Separate BDT classifiers are trained for each data-taking period and for the downstream and long categories.

The optimal BDT and PID cut values are determined separately for each subsample by optimising the figure of merit $\epsilon_{\rm sig}/\left(\frac{a}{2} + \sqrt{B}\right)$~\cite{Punzi:2003bu}, where $a=5$ quantifies the target level of significance in units of standard deviations ($\sigma$), $\epsilon_{\rm sig}$ is the efficiency of the signal selection determined from simulated events, and $B$ is the expected number of background events in the signal region, which is estimated by extrapolating the result of a fit to the invariant mass distribution of the data sidebands.
In the optimisation of the PID criteria, possible cross-feed backgrounds from misidentified decays to the other signal final states are also considered; their relative rates are obtained from data using the control modes containing $\Lc$ decays.
The optimised BDT requirements typically have signal efficiencies of around $50\,\%$ whilst rejecting over $90\,\%$ of the combinatorial background.
The optimised PID requirements have efficiencies around $60\,\%$ and reject over $95\,\%$ ($80\,\%$) of $\pion \rightarrow \kaon$ ($\kaon \rightarrow \pion$) cross-feed.
If more than one candidate is selected in any event, one is chosen at random and all others discarded -- this occurs in less than $2\,\%$ of selected events.

The efficiency of the selection requirements is studied using simulated events and, for the PID requirements, high-yield data control samples of $\Dz \to \Km \pip$ and $\Lz \to \proton \pim$ decays~\cite{LHCb-DP-2012-003}.
A multibody decay can in general proceed through intermediate states as well as through nonresonant amplitudes.
It is therefore necessary to model the variation of the efficiency, and to account for the distribution of signal events, over the phase space of the decay.
This is achieved, in a similar way as done for previous studies of $b$ baryon decays~\cite{LHCb-PAPER-2013-061,LHCb-PAPER-2013-056,LHCb-PAPER-2014-020}, by factorising the efficiency into a two-dimensional function of variables that describe the Dalitz plot~\cite{Dalitz:1953cp} and three one-dimensional functions for the angular variables.
Simulated events are binned in these variables in order to determine the selection efficiencies.
If no significant \bquark baryon signal is seen, the efficiency corresponding to a uniform phase-space distribution is used, and a systematic uncertainty is assigned to account for the variation across the phase space.
For modes with a significant yield, the distribution in the phase space is obtained with the \sPlot\ technique~\cite{Pivk:2004ty} with the \bquark baryon candidate invariant mass used as the control variable, and the efficiency corresponding to the observed distribution is used.

%% file: fit.tex
\section{Fit model and results}
\label{sec:Fit}

All signal and background yields, as well as the yields of $\Lb \to \Lc h^-$ decays, are determined using a single simultaneous unbinned extended maximum likelihood fit to the \bquark baryon candidate invariant mass distributions for each final state in the six subsamples, which correspond to the three data-taking periods and two reconstruction categories.
The probability density function (PDF) in each invariant mass distribution is defined as the sum of components accounting for signals, cross-feed contributions, combinatorial background and other backgrounds.
Fitting the subsamples simultaneously allows the use of common shape parameters, while fitting the different final states simultaneously facilitates the imposition of constraints on the level of cross-feed backgrounds.

Signal PDFs are known to have asymmetric tails that result from a combination of the effects of final-state radiation and stochastic tracking imperfections.
The signal mass distributions are each modelled by the sum of two Crystal Ball (CB) functions~\cite{Skwarnicki:1986xj} with a common mean and tails on opposite sides, where the high-mass tail accounts for non-Gaussian reconstruction effects.
The peak positions and overall widths of the CB functions are free parameters of the fit to data, while other shape parameters are determined from simulated samples, separately for each subsample, and are fixed in the fit to data.

Cross-feed backgrounds are also modelled by the sum of two CB functions.
The shape parameters are determined from simulation, separately for each subsample, and calibrated with the high-yield data control samples to account for the effects of the PID criteria.
In the fit to data, the misidentification rates are constrained to be consistent with expectation.

An exponential function is used to describe the combinatorial background, the yield of which is treated as independent for each subsample.
The shape parameter is taken to be the same for all data-taking periods, independently for each final state and reconstruction category.
In addition, components are included to account for possible backgrounds from \bquark baryon decays giving the same final state but with an extra soft (low energy) particle that is not reconstructed; examples include the photon that arises from $\PSigma^0 \to \Lz\gamma$ decay and the neutral pion in the $\Kstarp \to \Kp \piz$ decay.
Such partially reconstructed backgrounds are modelled by a generalised ARGUS function~\cite{Albrecht:1990cs} convolved with a Gaussian function, except in the case of the $\Lb \to \left(\Lz\pip\right)_{\Lc}\pim$ control mode where a nonparametric density estimate is used.
The shape parameters are determined from simulation, separately for the two reconstruction categories but for the data-taking periods combined, and are fixed in the fit to data; however, the yield of each partially reconstructed background is unconstrained in the fit.

In order to limit the number of free parameters in the fit, several additional constraints are imposed.
The yield of each cross-feed contribution is constrained within uncertainty to the yield of the corresponding correctly reconstructed decay multiplied by the appropriate misidentification rate.
The peak value of the signal shape is fixed to be the same for all \Lb decays, and the difference in peak values for \Xibz and \Lb decays is fixed to the known mass difference~\cite{LHCb-PAPER-2014-021}.
The widths of the signal shapes differ only between the two reconstruction categories, with a small correction factor, obtained from simulation, applied for the control channel modes with an intermediate \Lc decay.

In the $\Lz\Kp\Km$ final state, little or no background is expected in the \Xibz signal region.
Since likelihood fits cannot give reliable results if there are neither signal nor background candidates, the signal yields for $\Xibz\to\Lz\Kp\Km$ decays in the long reconstruction category are constrained to be non-negative.
All other signal yields are unconstrained.
The fit model and its stability are validated with ensembles of pseudoexperiments that are generated according to the fit model, with yields allowed to fluctuate around their expected values according to Poisson statistics.
No significant bias is found.

The results of the fit to data are given in Table~\ref{table:signalYields} and shown, for all subsamples combined, in Fig.~\ref{fig:dataFitsLc,pipi} for the $\Lb \to \left(\Lz\pip\right)_{\Lc}\pim$ control mode and the $\Lz\pip\pim$ signal final state, and in Fig.~\ref{fig:dataFitsKpi,KK} for the $\Lz\Kpm\pimp$ and $\Lz\Kp\Km$ signal final states.
The expected yield of misidentified $\Lb \to \Lz \pip \pim$ decays in the $\Lb \to \Lz \Kp \pim$ spectrum is $2.9 \pm 0.7$; that of $\Lb \to \Lz \Kp \pim$ decays in the $\Lb \to \Lz \Kp \Km$ spectrum is $3.2 \pm 0.5$; that of $\Lb \to \Lz \Kp \pim$ decays in the $\Lb \to \Lz \pip \pim$ spectrum is $14.0 \pm 2.0$; and that of $\Lb \to \Lz \Kp \Km$ decays in the $\Lb \to \Lz \Kp \pim$ spectrum is $35.3 \pm 2.8$.
All other cross-feed contributions are negligible.

The statistical significances of the $\Lb \to \Lz\pip\pim$, $\Lb \to \Lz\Kp\pim$, and $\Lb \to \Lz\Kp\Km$ decays, estimated from the change in log-likelihood between fits with and without these signal components, are $5.2\,\sigma$, $8.5\,\sigma$, and $20.5\,\sigma$ respectively.
The effects of systematic uncertainties on these values are given in Sec.~\ref{sec:bfresults}.
The statistical significances for all $\Xibz$ decays are less than $3\,\sigma$.

\begin{table}[tb]
\centering
\caption{\small
  Signal yields for the \Lb and \Xibz decay modes under investigation.
  The totals are simple sums and are not used in the analysis.
}
\label{table:signalYields}
\begin{tabular}{lccccc}
\hline
Mode & Run period & \multicolumn{4}{c}{Yield}\\
& & \multicolumn{2}{c}{\Lb} & \multicolumn{2}{c}{\Xibz}\\
& & downstream & long & downstream & long \\
\hline
              & 2011  & $10.2 \pm 5.5$ & $\phantom{1}8.7 \pm 4.7$ & $-0.6 \pm 2.4\phantom{-}$ & $4.9 \pm 3.2$\\
$\Lz\pip\pim$ & 2012a & $\phantom{1}9.1 \pm 5.2$ & $13.6 \pm 5.7$ & $5.3 \pm 3.6$ & $1.0 \pm 2.6$\\
              & 2012b & $17.2 \pm 7.1$ & $\phantom{1}6.2 \pm 4.6$ & $3.9 \pm 4.0$ & $4.1 \pm 2.7$\\
              & Total & \multicolumn{2}{c}{$65 \pm 14$} & \multicolumn{2}{c}{$19 \pm 8$} \\
\hline
              & 2011  & $20.9 \pm 6.4$ & $\phantom{1}8.2 \pm 3.5$ & $3.5 \pm 3.7$ & $-0.7 \pm 2.4\phantom{-}$\\
$\Lz\Kpm\pimp$& 2012a & $\phantom{1}9.3 \pm 3.7$ & $\phantom{1}1.7 \pm 3.6$ & $-0.1 \pm 1.7\phantom{-}$ & $0.3 \pm 1.5$\\
              & 2012b & $39.7 \pm 8.9$ & $16.9 \pm 5.1$ & $2.9 \pm 4.5$ & $-1.8 \pm 1.5\phantom{-}$\\
              & Total & \multicolumn{2}{c}{$97 \pm 14$} & \multicolumn{2}{c}{$4 \pm 7$} \\
\hline
              & 2011  & $32.3 \pm 6.4$ & $20.1 \pm 4.6$ & $0.6 \pm 2.3$ & $0.0 \pm 0.6$\\
$\Lz\Kp\Km$   & 2012a & $22.2 \pm 5.3$ & $15.9 \pm 4.2$ & $0.5 \pm 2.4$ & $0.0 \pm 0.5$\\
              & 2012b & $60.5 \pm 8.5$ & $34.4 \pm 6.1$ & $3.0 \pm 2.7$ & $0.0 \pm 0.6$\\
              & Total & \multicolumn{2}{c}{$185 \pm 15$} & \multicolumn{2}{c}{$4 \pm 4$ } \\
\hline
                                 & 2011  & $78.1 \pm 9.1$ & $78.9 \pm 9.2$ \\
$\left(\Lz\pip\right)_{\Lc}\pim$ & 2012a & $45.0 \pm 7.0$ & $63.0 \pm 8.3$ \\
                                 & 2012b & $115.3 \pm 11.1$ & $90.7 \pm 9.8$ \\
              & Total & \multicolumn{2}{c}{$471 \pm 22$} \\
\hline
\end{tabular}
\end{table}

\begin{figure}[!tb]
\centering
\includegraphics*[width=0.49\textwidth]{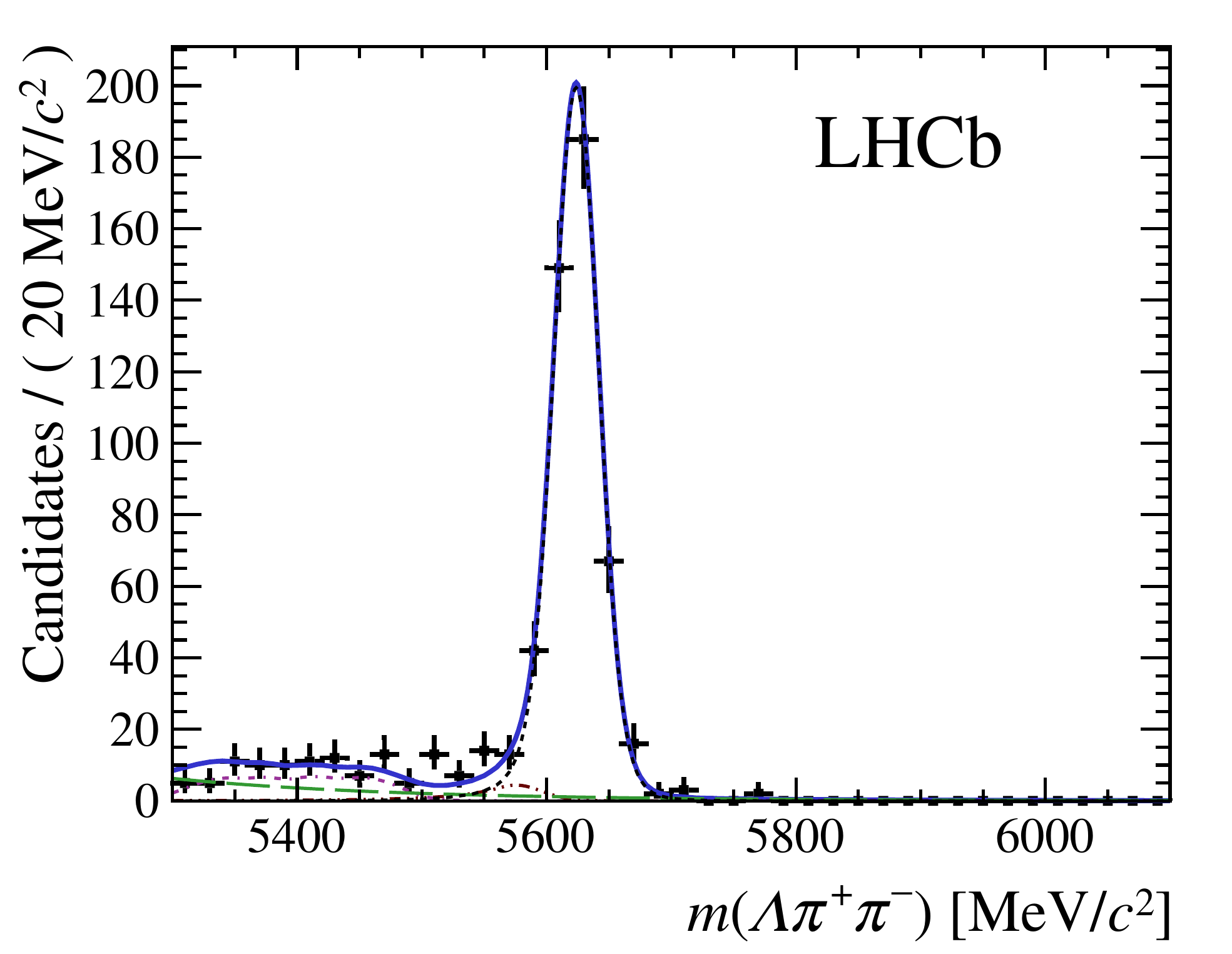}
\includegraphics*[width=0.49\textwidth]{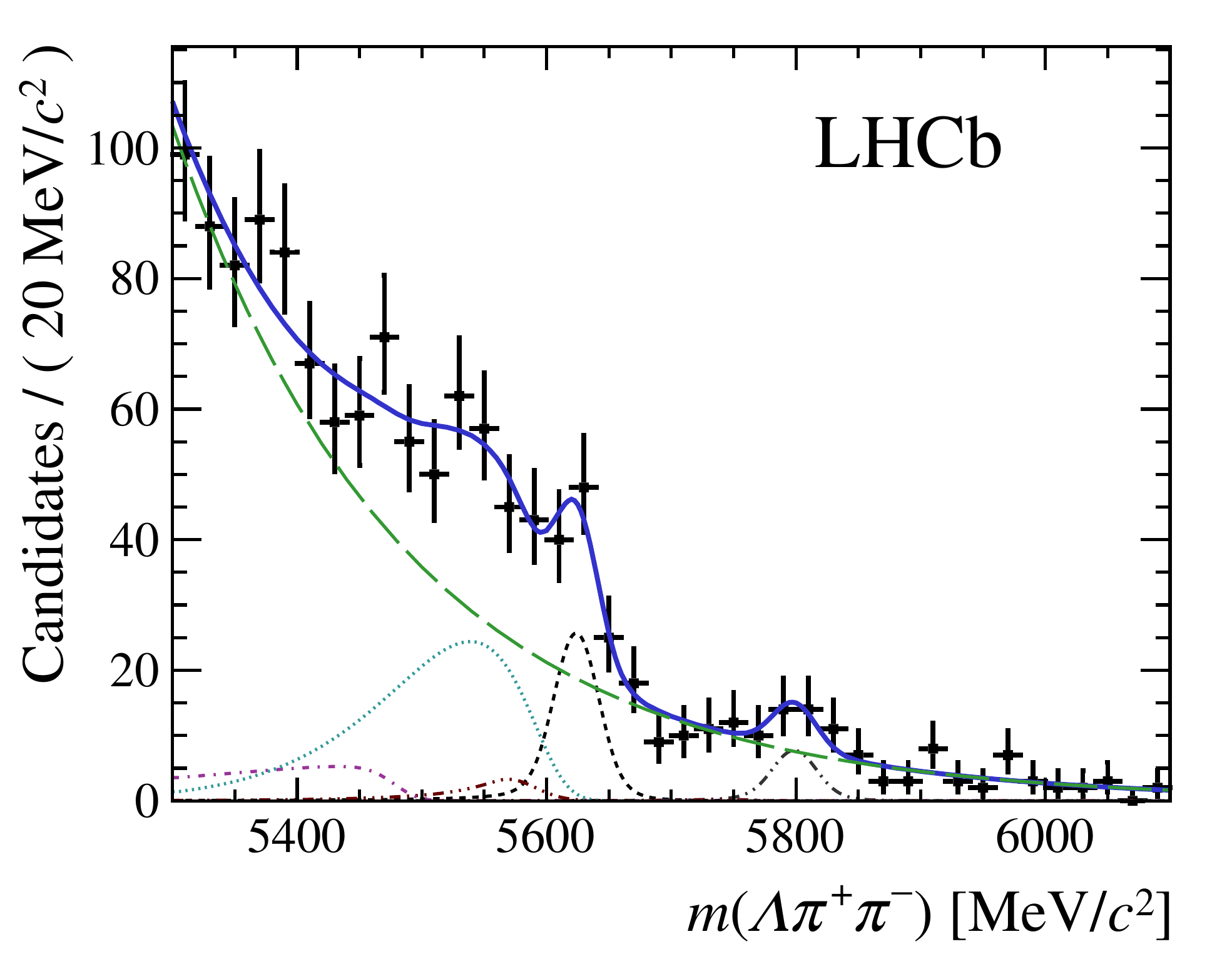}\\
\caption{\small
  Results of the fit for the (left) $\Lb \to \left(\Lz\pip\right)_{\Lc}\pim$ control mode and (right) $\Lz\pip\pim$ signal final states, for all subsamples combined.
  Superimposed on the data are the total result of the fit as a solid blue line, the \Lb (\Xibz) decay as a short-dashed black (double dot-dashed grey) line, cross-feed as triple dot-dashed brown lines, the combinatorial background as a long-dashed green line, and partially reconstructed background components with either a missing neutral pion as a dot-dashed purple line or a missing soft photon as a dotted cyan line.
}
\label{fig:dataFitsLc,pipi}
\end{figure}

\begin{figure}[!tb]
\centering
\includegraphics*[width=0.49\textwidth]{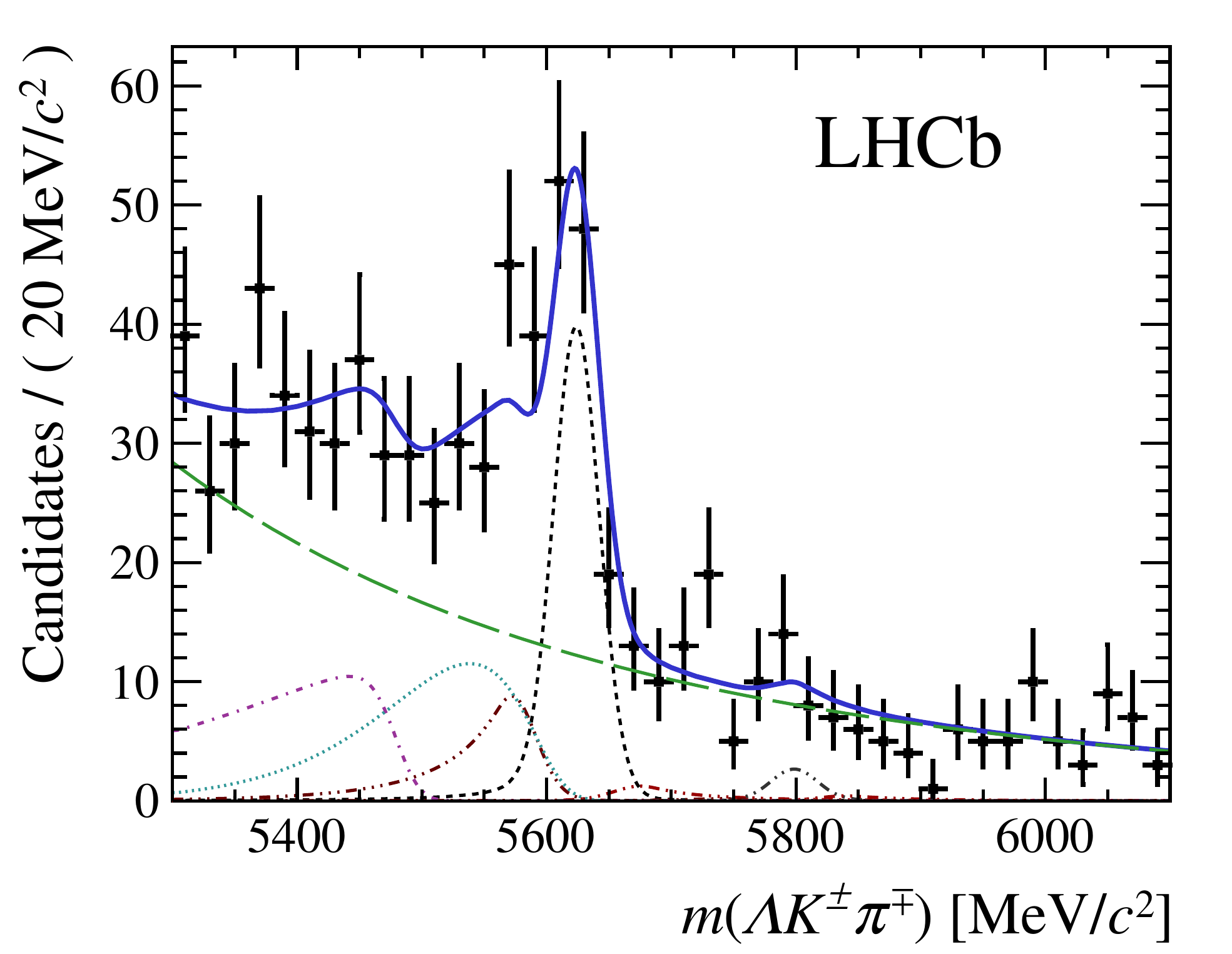}
\includegraphics*[width=0.49\textwidth]{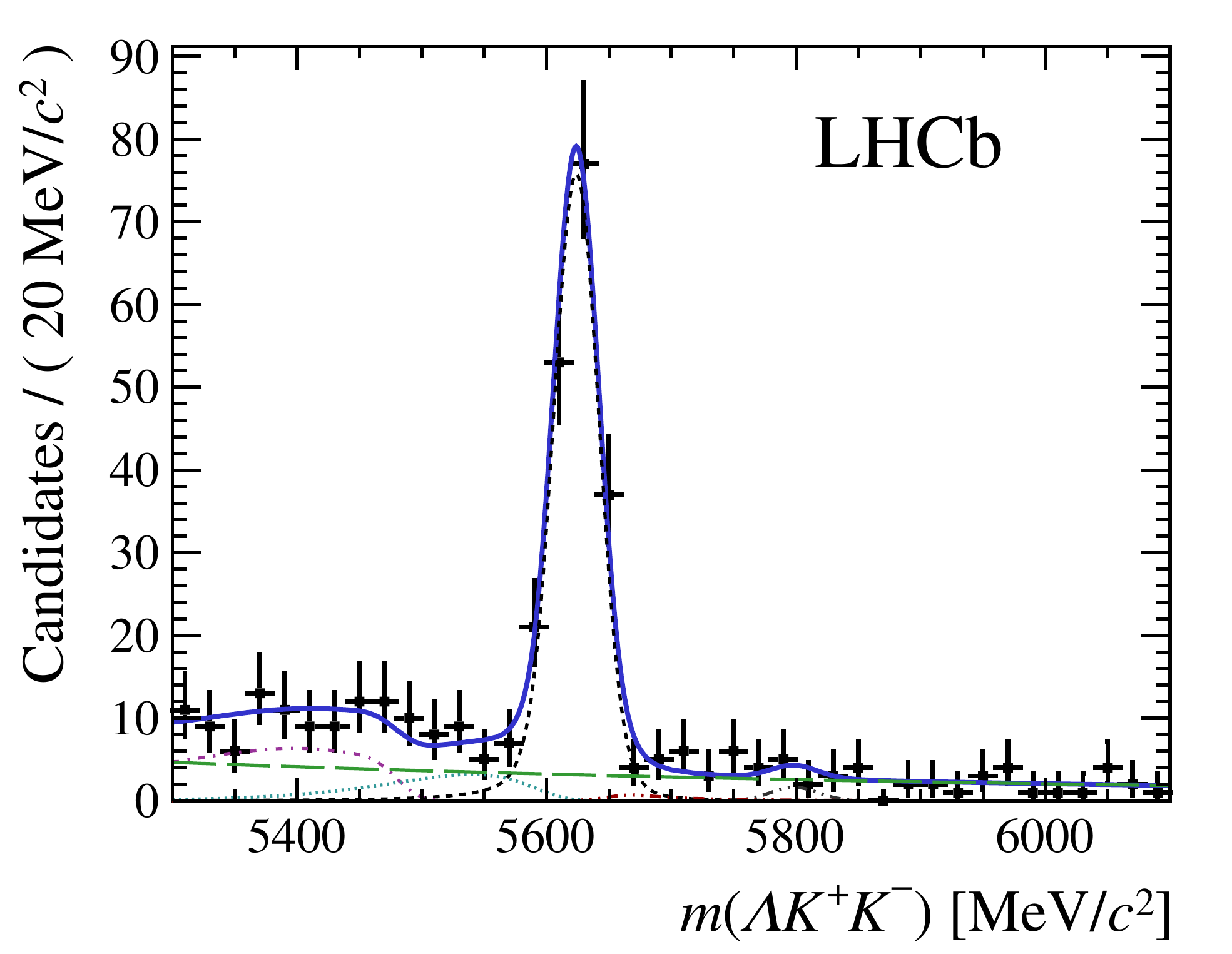}\\
\caption{\small
  Results of the fit for the (left) $\Lz\Kpm\pimp$ and (right) $\Lz\Kp\Km$ final states, for all subsamples combined.
  Superimposed on the data are the total result of the fit as a solid blue line, the \Lb (\Xibz) decay as a short-dashed black (double dot-dashed grey) line, cross-feed as triple dot-dashed brown lines, the combinatorial background as a long-dashed green line, and partially reconstructed background components with either a missing neutral pion as a dot-dashed purple line or a missing soft photon as a dotted cyan line.
}
\label{fig:dataFitsKpi,KK}
\end{figure}

\begin{figure}[tb]
\centering
\includegraphics*[width=0.49\textwidth]{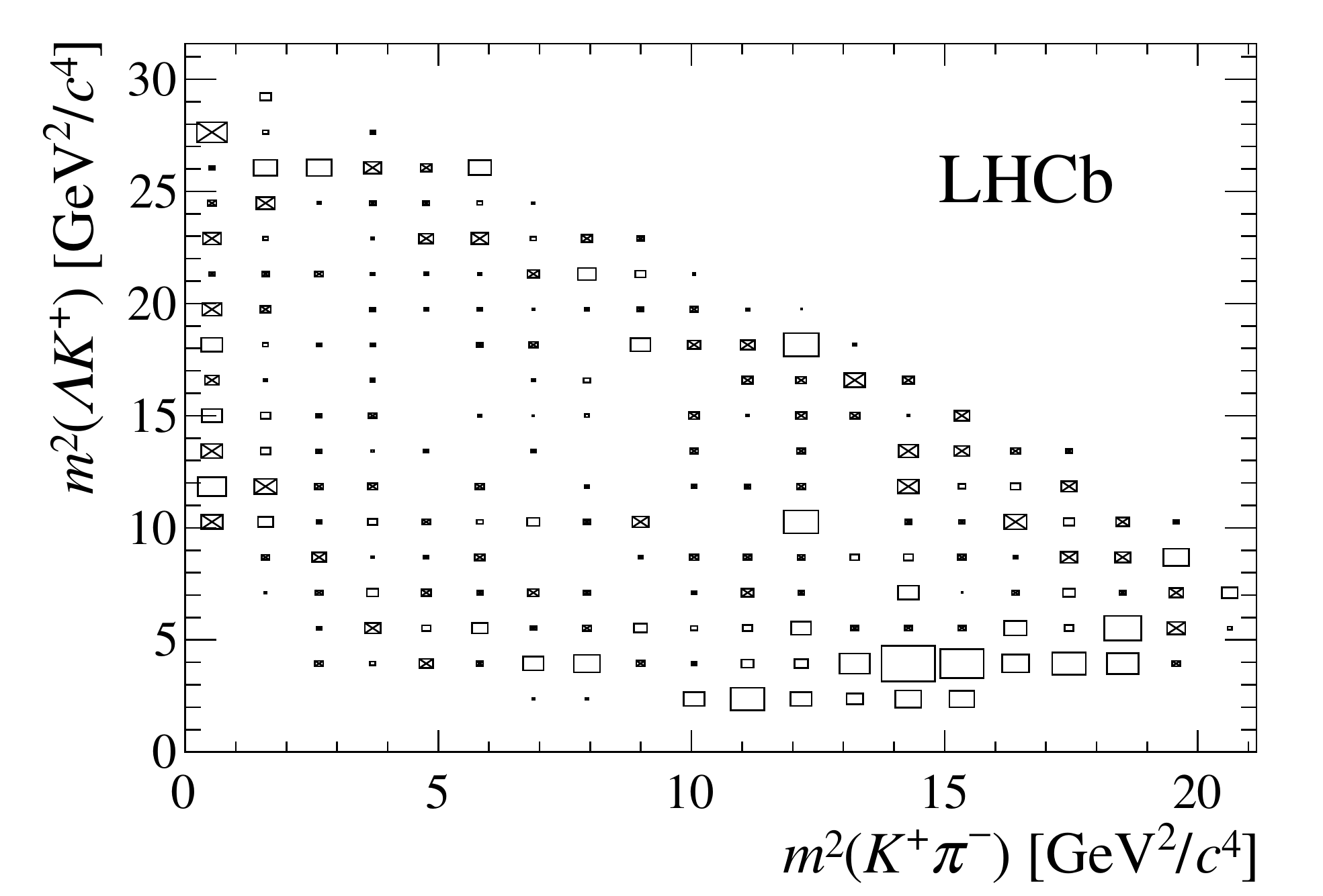}
\includegraphics*[width=0.49\textwidth]{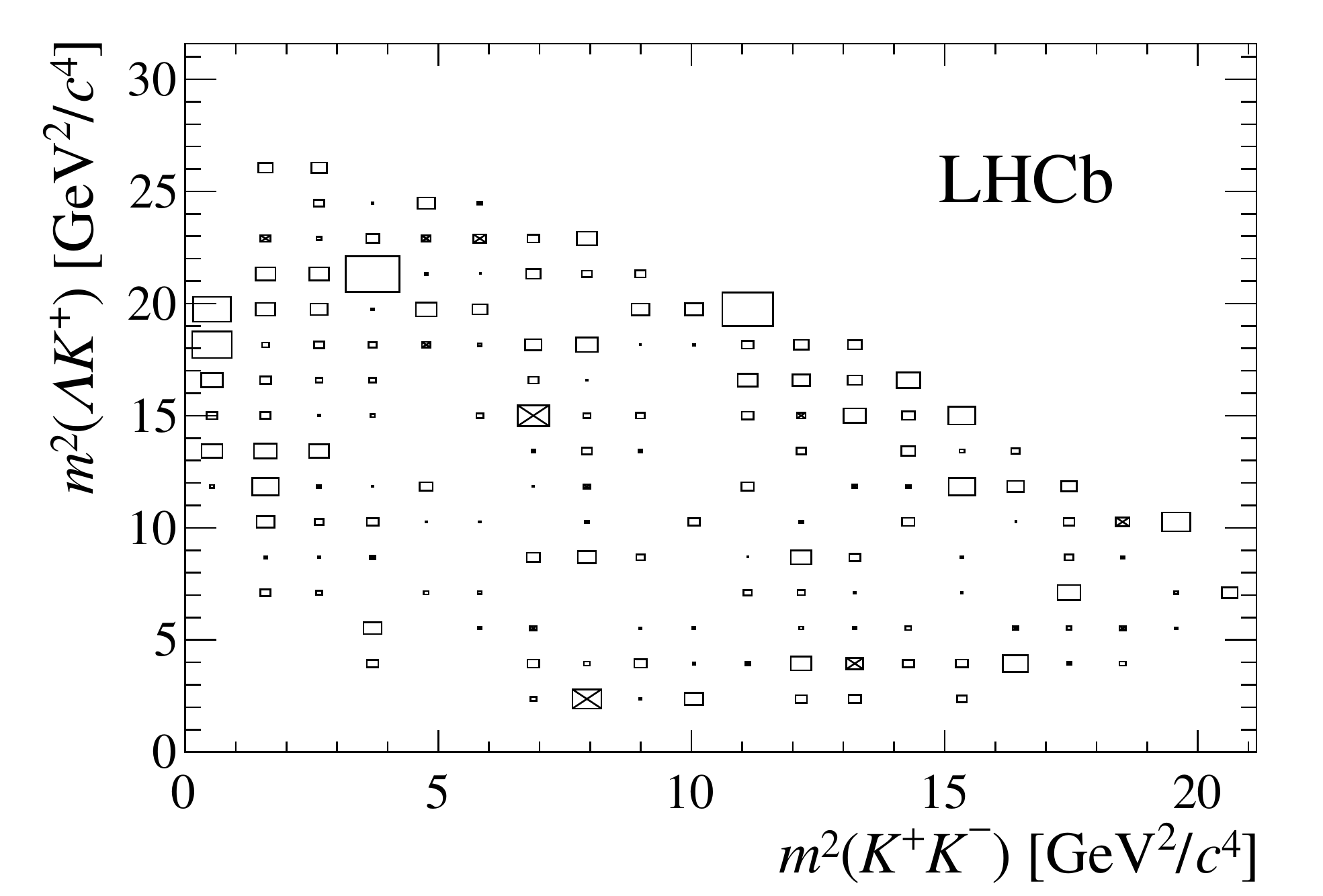}
\caption{\small
  Background-subtracted and efficiency-corrected Dalitz plot distributions for (left) $\Lb \to \Lz\Kp\pim$ and (right) $\Lb \to \Lz\Kp\Km$ with data from all subsamples combined. Boxes with a cross indicate negative values.
}
\label{fig:dalitz_plot}
\end{figure}

As significant yields are obtained for $\Lb \to \Lz\Kp\pim$ and $\Lb \to \Lz\Kp\Km$ decays, their Dalitz plot distributions are obtained from data using the \sPlot\ technique and applying event-by-event efficiency corrections based on the position of the decay in the phase space.
These distributions are used to determine the average efficiencies for these channels, and are shown in Fig.~\ref{fig:dalitz_plot}, where the negative (crossed) bins occur due to the statistical nature of the background subtraction.
The  $\Lb \to \Lz\Kp\Km$ signal seen at low $m^2(\Kp\Km)$ is consistent with the recent observation of the $\Lb \to \Lz\phi$ decay~\cite{LHCb-PAPER-2016-002}.
Although the statistical significance of the $\Lb \to \Lz\pip\pim$ channel is over $5\,\sigma$, the uncertainty on its Dalitz plot distribution is too large for this method of determining the average efficiency to be viable.

%% file: systematics.tex
\section{Systematic uncertainties}
\label{sec:Systematics}

Systematic uncertainties in the branching fraction measurements are minimised by the choice of a normalisation channel with similar topology and final-state particles.
There are residual uncertainties due to approximations made in the fit model, imperfect knowledge of the efficiency, and the uncertainty on the normalisation channel yield.
The systematic uncertainties are evaluated separately for each subsample, with correlations taken into account in the combination of results.
A summary of the uncertainties assigned on the combined results is given in Table~\ref{tab:systematics}.

The systematic uncertainty from the fit model is evaluated by using alternative shapes for each of the components, for both the charmless and \Lc spectra.
The double Crystal Ball function used for the signal component is replaced with the sum of two Gaussian functions with a common mean.
The partially reconstructed background shapes are replaced with nonparametric functions determined from simulation.
The combinatorial background model is changed from an exponential function to a second-order polynomial shape.
In addition, the effect of varying fixed parameters of the model within their uncertainties is evaluated with pseudoexperiments and added in quadrature to the fit model systematic uncertainty.

There are several sources of systematic uncertainty related to the evaluation of the relative efficiency.
The first is due to the finite size of the simulation samples, and is determined from the effect of fluctuating the efficiency, within uncertainties, in each phase-space bin.
The second is determined from the variation of the efficiency across the phase space, and is relevant only for modes without a significant signal yield.
The third, from the uncertainty on the kinematical agreement between the signal mode and the PID control modes, is determined by varying the binning of these control samples.
Finally, the effects of the vetoes applied to remove charmed intermediate states are investigated by studying the variation in the result with different requirements.

In order to determine relative branching fractions, it is necessary to account also for the statistical uncertainty in the yield of the $\Lb \to \left(\Lz\pip\right)_{\Lc}\pim$ normalisation channel.
The uncertainty on its branching fraction is included when converting results to absolute branching fractions.
The total systematic uncertainty is determined as the sum in quadrature of all contributions.

\begin{table}[!tb]
\centering
\caption{\small
  Systematic uncertainties (in units of $10^{-3}$) on the branching fraction ratios reported in Sec.~\ref{sec:bfresults}.
  The total is the sum in quadrature of all contributions.
}
\label{tab:systematics}
\begin{tabular}{lccccccc}
\hline
            & Fit   & Efficiency & Phase space  & PID   & Vetoes & $\Lc\pim$ yield & Total \\
\hline
$\Lb\to\Lz\pip\pim$   & $8.4$ & $2.0$  & $19.7\phantom{0}$ & $0.4$ & $2.2$ & $3.5$  & $21.9\phantom{0}$ \\
$\Lb\to\Lz\Kp\pim$    & $1.7$ & $11.7\phantom{0}$ &  ---   & $2.9$ & $1.3$ & $4.6$  & $13.1\phantom{0}$ \\
$\Lb\to\Lz\Kp\Km$     & $6.7$ & $5.4$  &  ---   & $4.2$ & $2.2$ & $15.9\phantom{0}$ & $18.7\phantom{0}$ \\
$\Xibz\to\Lz\pip\pim$ & $4.1$ & $0.7$  & $7.0$  & $0.1$ &  ---  & $1.2$  & $8.2$ \\
$\Xibz\to\Lz\pip\Km$  & $1.5$ & $0.4$  & $3.5$  & $0.1$ &  ---  & $0.7$  & $4.0$ \\
$\Xibz\to\Lz\Kp\Km$   & $0.1$ & $0.1$  & $0.8$  & $0.0$ &  ---  & $0.2$  & $0.8$ \\
\hline
\end{tabular}
\end{table}

%% file: bfresult.tex
\section{Branching fraction results}
\label{sec:bfresults}

The relative branching fractions for the \Lb decay modes are determined according to
\begin{equation}
  \label{eq:partialBF}
  \frac{{\cal B}(\Lb \to \Lz h^+h^{\prime -})}{{\cal B}(\Lb \to \left( \Lz\pip \right)_{\Lc} \pim)} =
  \frac{N(\Lb \to \Lz h^+h^{\prime -})}{N(\Lb \to \left( \Lz\pip \right)_{\Lc} \pim)} \times
  \frac{\epsilon(\Lb \to \left( \Lz\pip \right)_{\Lc} \pim)}{\epsilon(\Lb \to \Lz h^+h^{\prime -})} \, ,
\end{equation}
where $N$ denotes the yield determined from the maximum likelihood fit to data, as described in Sec.~\ref{sec:Fit}, and $\epsilon$ denotes the efficiency, as described in Sec.~\ref{sec:Selection}.
For the \Xibz decay modes the expression is modified to account for  the fragmentation fractions $f_{\Xibz}$ and $f_{\Lb}$, \ie\ the probability that a $b$ quark hadronises into either a \Xibz or \Lb baryon,
\begin{equation}
  \label{eq:partialBF-Xib}
  \frac{f_{\Xibz}}{f_{\Lb}} \times
  \frac{{\cal B}(\Xibz \to \Lz h^+h^{\prime -})}{{\cal B}(\Lb \to \left( \Lz\pip \right)_{\Lc} \pim)} =
  \frac{N(\Xibz \to \Lz h^+h^{\prime -})}{N(\Lb \to \left( \Lz\pip \right)_{\Lc} \pim)} \times
  \frac{\epsilon(\Lb \to \left( \Lz\pip \right)_{\Lc} \pim)}{\epsilon(\Xibz \to \Lz h^+h^{\prime -})} \, .
\end{equation}
Since $f_{\Xibz}$ is yet to be measured, the product of quantities on the left-hand side of Eq.~(\ref{eq:partialBF-Xib}) is reported.

The ratios in Eq.~(\ref{eq:partialBF}) and Eq.~(\ref{eq:partialBF-Xib}) are determined separately for each subsample, and the independent measurements of each quantity are found to be consistent.
The results for the subsamples are then combined, taking correlations among the systematic uncertainties into account, giving
\begin{equation*}
  \begin{array}{rcr}
    \mathlarger{\frac{{\cal B}(\Lb\to\Lz\pip\pim)}{{\cal B}(\Lb \to \left( \Lz\pip \right)_{\Lc} \pim)}} & = & \left(7.3 \pm 1.9 \pm 2.2 \right) \times 10^{-2}\,, \\ [2.5ex]
    \mathlarger{\frac{{\cal B}(\Lb\to\Lz\Kp\pim)}{{\cal B}(\Lb \to \left( \Lz\pip \right)_{\Lc} \pim)}} & = & \left(8.9 \pm 1.2 \pm 1.3 \right) \times 10^{-2}\,, \\ [2.5ex]
    \mathlarger{\frac{{\cal B}(\Lb\to\Lz\Kp\Km)}{{\cal B}(\Lb \to \left( \Lz\pip \right)_{\Lc} \pim)}} & = & \left(25.3 \pm 1.9 \pm 1.9 \right) \times 10^{-2}\,, \\ [2.5ex]
    \mathlarger{\frac{f_{\Xibz}}{f_{\Lb}} \times \frac{{\cal B}(\Xibz\to\Lz\pip\pim)}{{\cal B}(\Lb \to \left( \Lz\pip \right)_{\Lc} \pim)}} & = & \left(2.0 \pm 1.0 \pm 0.8 \right) \times 10^{-2}\,, \\ [2.5ex]
    \mathlarger{\frac{f_{\Xibz}}{f_{\Lb}} \times \frac{{\cal B}(\Xibz\to\Lz\Km\pip)}{{\cal B}(\Lb \to \left( \Lz\pip \right)_{\Lc} \pim)}} & = & \left({-}0.1 \pm 0.8 \pm 0.4 \right) \times 10^{-2}\,,
  \end{array}
\end{equation*}
where the first quoted uncertainty is statistical and the second is systematic.
The significances for the $\Lb \to \Lz\pip\pim$, $\Lb \to \Lz\Kp\pim$, and $\Lb \to \Lz\Kp\Km$ modes, including the effects of systematic uncertainties on the yields, are $4.7\,\sigma$, $8.1\,\sigma$, and $15.8\,\sigma$ respectively.
These are calculated from the change in log-likelihood, after the likelihood obtained from the fit is convolved with a Gaussian function with width corresponding to the systematic uncertainty on the yield.

The relative branching fractions are multiplied by ${\cal B}(\Lb \to \left( \Lz\pip \right)_{\Lc} \pim)$ to obtain absolute branching fractions.
The normalisation channel product branching fraction is evaluated to be $(6.29 \pm 0.78)\times10^{-5}$ from measurements of ${\cal B}(\Lb \to \Lc \pim)$~\cite{LHCb-PAPER-2014-004}, ${\cal B}(\Lc \to \Lz\pip)/{\cal B}(\Lc \to \proton\Km\pip)$~\cite{Link:2005ut} and ${\cal B}(\Lc \to \proton\Km\pip)$~\cite{Zupanc:2013iki}.

As the likelihood function for $\Xibz\to\Lz\Kp\Km$ decays is not reliable, owing to the absence of data in the signal region in the long reconstruction category, a Bayesian approach~\cite{LHCb-PAPER-2012-025} is used to obtain an upper limit on the branching fraction of this decay mode.
The \Xibz signal region, $5763 < m(\Lz h^+ h^-) < 5823 \mevcc$, is assumed to contain the Poisson distributed sum of background and signal components.
The prior probability distribution for the signal rate is flat, whereas the prior for the background rate is a Gaussian distribution based on the expectation from the maximum likelihood fit, found by extrapolating the combinatorial background component from the fit into the signal region.
Both of these prior distributions are truncated to remove the unphysical (negative) region.
Log-normal priors are used for the normalisation mode yield, the signal and normalisation channel efficiencies, and all other sources of systematic uncertainty.
The posterior probability distribution is obtained by integrating over the nuisance parameters using Markov chain Monte Carlo~\cite{gilks1995markov}.
For consistency, the same method is used to obtain upper limits on the branching fractions of all modes which do not have significant yields.

The results for the absolute branching fractions are
\begin{equation*}
  \begin{array}{rcr}
    {\cal B}(\Lb\to\Lz\pip\pim) & = & \left(4.6 \pm 1.2 \pm 1.4 \pm 0.6 \right) \times 10^{-6}\,, \\ [0.8ex]
    {\cal B}(\Lb\to\Lz\Kp\pim) & = & \left(5.6 \pm 0.8 \pm 0.8 \pm 0.7 \right) \times 10^{-6}\,, \\  [0.8ex]
    {\cal B}(\Lb\to\Lz\Kp\Km) & = & \left(15.9 \pm 1.2 \pm 1.2 \pm 2.0 \right) \times 10^{-6}\,, \\
    \frac{f_{\Xibz}}{f_{\Lb}} \times {\cal B}(\Xibz\to\Lz\pip\pim) & = & \left(1.3 \pm 0.6 \pm 0.5 \pm 0.2 \right) \times 10^{-6}\,, \\
    & < & \multicolumn{1}{l}{~1.7 ~(2.1) \times 10^{-6} \ \text{at 90~(95)\,\% confidence level} \, ,} \\
    \frac{f_{\Xibz}}{f_{\Lb}} \times {\cal B}(\Xibz\to\Lz\Km\pip) & = & \left({-}0.6 \pm 0.5 \pm 0.3 \pm 0.1 \right) \times 10^{-6}\,, \\
    & < & \multicolumn{1}{l}{~0.8 ~(1.0) \times 10^{-6} \ \text{at 90~(95)\,\% confidence level} \, ,} \\
    \frac{f_{\Xibz}}{f_{\Lb}} \times {\cal B}(\Xibz\to\Lz\Kp\Km) & < & \multicolumn{1}{l}{~0.3 ~(0.4) \times 10^{-6} \ \text{at 90~(95)\,\% confidence level} \, ,}
  \end{array}
\end{equation*}
where the last quoted uncertainty is due to the precision with which the normalisation channel branching fraction is known.

%% file: acp.tex
\section{\boldmath \CP asymmetry measurements}
\label{sec:acp}

The significant yields observed for the $\Lb\to \Lz\Kp\pim$ and $\Lz\Kp\Km$ decays allow measurements of their phase-space integrated \CP asymmetries.
The simultaneous extended maximum likelihood fit is modified to allow the determination of the raw asymmetry, defined as
\begin{equation}
  \label{eq:ACP_master}
  {\cal A}^{\rm raw}_{\CP} = \frac{N^{\rm corr}_{f} - N^{\rm corr}_{\bar{f}}}{N^{\rm corr}_{f} + N^{\rm corr}_{\bar{f}}}\,,
\end{equation}
where $N^{\rm corr}_{f}$ ($N^{\rm corr}_{\bar{f}}$) is the efficiency-corrected yield for $\Lb$ ($\Lbbar$) decays.
The use of the efficiency-corrected yields accounts for the possibility that there may be larger \CP violation effects in certain regions of phase space, as seen in other charmless three-body \bquark~hadron decays~\cite{LHCb-PAPER-2014-044}.

To measure the parameter of the underlying \CP violation, the raw asymmetry has to be corrected for possible small detection (${\cal A}_{\rm D}$) and production (${\cal A}_{\rm P}$) asymmetries,
${\cal A}_{\CP} = {\cal A}^{\rm raw}_{\CP} - \left({\cal A}_{\rm P} + {\cal A}_{\rm D}\right)$.
This can be conveniently achieved with the $\Lb \to \left( \Lz\pip \right)_{\Lc} \pim$ control mode, which is expected to have negligible \CP violation.
Since this mode shares the same initial state as the decay of interest, it has the same production asymmetry; moreover, the final-state selection differs only in the PID requirements and therefore most detection asymmetry effects also cancel.
Thus,
\begin{equation}
  {\cal A}_{\CP}(\Lb\to \Lz h^+h^{\prime -}) = {\cal A}^{\rm raw}_{\CP}(\Lb\to \Lz h^+h^{\prime -}) - {\cal A}^{\rm raw}_{\CP}(\Lb \to \left( \Lz\pip \right)_{\Lc} \pim) \, .
\end{equation}

The measured raw asymmetries, including the efficiency correction for the signal modes, for $\Lb\to \Lz\Kp\pim$, $\Lb\to \Lz\Kp\Km$, and $\Lb \to \left( \Lz\pip \right)_{\Lc} \pim$ are determined by performing the fit with the data separated into \Lb or \Lbbar candidates, depending on the charge of the \proton from the $\Lz \to \proton\pim$ decay.
They are found to be
${\cal A}^{\rm raw}_{\CP}(\Lb\to \Lz\Kp\pim) = -0.46 \pm 0.23$,
${\cal A}^{\rm raw}_{\CP}(\Lb\to \Lz\Kp\Km) = -0.21 \pm 0.10$ and
${\cal A}^{\rm raw}_{\CP}(\Lb \to \left( \Lz\pip \right)_{\Lc} \pim) = 0.07 \pm 0.07$, where the uncertainties are statistical only.
The asymmetries for the background components are found to be consistent with zero, as expected.

\begin{table}[!tb]
\centering
\caption{\small
  Systematic uncertainties on ${\cal A}_{\CP}$ (in units of $10^{-3}$).}
\label{table:acpSystSummary}
\begin{tabular}{lcc}
\hline \\ [-2.4ex]
& ${\cal A}_{\CP}(\Lb\to \Lz\Kp\pim)$ & ${\cal A}_{\CP}(\Lb\to\Lz\Kp\Km)$ \\
\hline
Control mode & $\phantom{1}66$ & $57$ \\
PID asymmetry & $\phantom{1}20$ & -- \\
Fit model & $\phantom{1}27$ & $32$  \\
Fit bias & $\phantom{1}14$ & $\phantom{0}4$ \\
Efficiency uncertainty & $\phantom{1}80$ & $28$  \\
\hline
Total & $110$ & $71$ \\
\hline
\end{tabular}
\end{table}

Several sources of systematic uncertainty are considered, as summarised in Table~\ref{table:acpSystSummary}.
The uncertainty on ${\cal A}_{\rm P} + {\cal A}_{\rm D}$ comes directly from the result of the fit to $\Lb \to \left( \Lz\pip \right)_{\Lc} \pim$ decays.
The effect of variations of the detection asymmetry with the decay kinematics, which can be slightly different for reconstructed signal and control modes, is negligible.
However, for the $\Lb\to \Lz\Kp\pim$ channel, a possible asymmetry in kaon detection, which is taken to be $2\,\%$~\cite{LHCb-PAPER-2014-013}, has to be accounted for.
Effects related to the choices of signal and background models, possible intrinsic fit biases, and uncertainties in the efficiencies are evaluated in a similar way as for the branching fraction measurements.
The total systematic uncertainty is obtained by summing all contributions in quadrature.

The results for the phase-space integrated \CP asymmetries, with correlations taken into account, are
\begin{eqnarray*}
  {\cal A}_{\CP}(\Lb\to \Lz\Kp\pim) & = & -0.53 \pm 0.23 \pm 0.11 \, ,\\
  {\cal A}_{\CP}(\Lb\to \Lz\Kp\Km)  & = & -0.28 \pm 0.10 \pm 0.07 \, ,
\end{eqnarray*}
where the uncertainties are statistical and systematic, respectively.
These are both less than $3\,\sigma$ from zero, indicating consistency with \CP symmetry.

%% file: conclusions.tex
\section{Conclusions}
\label{sec:Conclusions}

Using a data sample collected by the LHCb experiment corresponding to an integrated luminosity of $3\invfb$ of high-energy $pp$ collisions, a search for charmless three-body decays of \bquark baryons to the $\Lz \pip\pim$, $\Lz \Kpm\pimp$ and $\Lz \Kp\Km$ final states has been performed.
The $\Lb \to \Lz\Kp\pim$ and $\Lb \to \Lz\Kp\Km$ decay modes are observed for the first time, and their branching fractions and \CP asymmetry parameters are measured.
No evidence is seen for \CP asymmetry in the phase-space integrated decay rates of these modes.
Evidence is seen for the $\Lb \to \Lz\pip\pim$ decay, with a branching fraction somewhat larger than predicted by theoretical calculations~\cite{Guo:1998eg,Arunagiri:2003gu,Leitner:2006nb}, and limits are set on the branching fractions of $\Xibz \to \Lz\pip\pim$, $\Xibz \to \Lz\Km\pip$, and $\Xibz \to \Lz\Kp\Km$ decays.
These results motivate further studies, both experimental and theoretical, into \Lb and \Xibz decays to $\Lz h^+ h^{\prime -}$ final states.

%% file: acknowledgements.tex
\section*{Acknowledgements}

\noindent We express our gratitude to our colleagues in the CERN
accelerator departments for the excellent performance of the LHC. We
thank the technical and administrative staff at the LHCb
institutes. We acknowledge support from CERN and from the national
agencies: CAPES, CNPq, FAPERJ and FINEP (Brazil); NSFC (China);
CNRS/IN2P3 (France); BMBF, DFG and MPG (Germany); INFN (Italy); 
FOM and NWO (The Netherlands); MNiSW and NCN (Poland); MEN/IFA (Romania); 
MinES and FANO (Russia); MinECo (Spain); SNSF and SER (Switzerland); 
NASU (Ukraine); STFC (United Kingdom); NSF (USA).
We acknowledge the computing resources that are provided by CERN, IN2P3 (France), KIT and DESY (Germany), INFN (Italy), SURF (The Netherlands), PIC (Spain), GridPP (United Kingdom), RRCKI and Yandex LLC (Russia), CSCS (Switzerland), IFIN-HH (Romania), CBPF (Brazil), PL-GRID (Poland) and OSC (USA). We are indebted to the communities behind the multiple open 
source software packages on which we depend.
Individual groups or members have received support from AvH Foundation (Germany),
EPLANET, Marie Sk\l{}odowska-Curie Actions and ERC (European Union), 
Conseil G\'{e}n\'{e}ral de Haute-Savoie, Labex ENIGMASS and OCEVU, 
R\'{e}gion Auvergne (France), RFBR and Yandex LLC (Russia), GVA, XuntaGal and GENCAT (Spain), Herchel Smith Fund, The Royal Society, Royal Commission for the Exhibition of 1851 and the Leverhulme Trust (United Kingdom).

%% file: LHCb_HD_authorlist_2016-01-19.tex
\centerline{\large\bf LHCb collaboration}
\begin{flushleft}
\small
R.~Aaij$^{39}$, 
C.~Abell\'{a}n~Beteta$^{41}$, 
B.~Adeva$^{38}$, 
M.~Adinolfi$^{47}$, 
Z.~Ajaltouni$^{5}$, 
S.~Akar$^{6}$, 
J.~Albrecht$^{10}$, 
F.~Alessio$^{39}$, 
M.~Alexander$^{52}$, 
S.~Ali$^{42}$
% ...

{\footnotesize \it
% ...
$ ^{5}$Clermont Universit\'{e}, Universit\'{e} Blaise Pascal, CNRS/IN2P3, LPC, Clermont-Ferrand, France\\
$ ^{6}$CPPM, Aix-Marseille Universit\'{e}, CNRS/IN2P3, Marseille, France\\
% ...
$ ^{10}$Fakult\"{a}t Physik, Technische Universit\"{a}t Dortmund, Dortmund, Germany\\
% ...
$ ^{38}$Universidad de Santiago de Compostela, Santiago de Compostela, Spain\\
$ ^{39}$European Organization for Nuclear Research (CERN), Geneva, Switzerland\\
% ...
$ ^{41}$Physik-Institut, Universit\"{a}t Z\"{u}rich, Z\"{u}rich, Switzerland\\
$ ^{42}$Nikhef National Institute for Subatomic Physics, Amsterdam, The Netherlands\\
% ...
$ ^{47}$H.H. Wills Physics Laboratory, University of Bristol, Bristol, United Kingdom\\
% ...
$ ^{52}$School of Physics and Astronomy, University of Glasgow, Glasgow, United Kingdom\\
% ...
}
\end{flushleft}